\pdfoutput=1

\documentclass[11pt]{article}

\usepackage[]{emnlp2021}

\usepackage{times}
\usepackage{latexsym}

\usepackage[T1]{fontenc}

\usepackage[utf8]{inputenc}

\usepackage{times}
\usepackage{latexsym}
\usepackage{subcaption}
\usepackage{adjustbox}
\usepackage{multirow}
\usepackage{graphics}
\usepackage{microtype}
\usepackage{amsmath}
\usepackage{amssymb}
\usepackage{bm}
\usepackage{graphicx}
\usepackage{xcolor}
\usepackage{array}

\title{Unsupervised Corpus Aware Language Model Pre-training\\ for Dense Passage Retrieval}

\author{Luyu Gao and Jamie Callan \\
  Language Technologies Institute \\
  Carnegie Mellon University \\
  \texttt{\{luyug, callan\}@cs.cmu.edu}  }

\begin{document}
\maketitle
\begin{abstract}
Recent research demonstrates the effectiveness of using fine-tuned language models~(LM) for dense retrieval. However, dense retrievers are hard to train, typically requiring heavily engineered fine-tuning pipelines to realize their full potential. In this paper, we identify and address two underlying problems of dense retrievers: i)~fragility to training data noise and ii)~requiring large batches to robustly learn the embedding space. We use the recently proposed Condenser pre-training architecture, which learns to condense information into the dense vector through LM pre-training. On top of it, we propose coCondenser, which adds an unsupervised corpus-level contrastive loss to warm up the passage embedding space. Retrieval experiments on MS-MARCO, Natural Question, and Trivia QA datasets show that coCondenser removes the need for heavy data engineering such as augmentation, synthesis, or filtering, as well as the need for large batch training. It shows comparable performance to RocketQA, a state-of-the-art, heavily engineered system, using simple small batch fine-tuning.\footnote{Our code is available at \url{https://github.com/luyug/Condenser}}
\end{abstract}
\section{Introduction}
 Building upon the advancements of pre-trained language models~(LM; \citet{devlin-etal-2019-bert, Liu2019RoBERTaAR}), dense retrieval has become an effective paradigm for text retrieval~\cite{lee-etal-2019-latent,chang2020pretraining,karpukhin-etal-2020-dense,Qu2020RocketQAAO}. Recent research has however found that fine-tuning dense retrievers to realize their capacity requires carefully designed fine-tuning techniques. Early works include iterative negative mining~\cite{xiong2021approximate} and multi-vector representations~\cite{Luan2020SparseDA}. The recent RocketQA system~\cite{Qu2020RocketQAAO} significantly improves the performance of a dense retriever by designing an optimized fine-tuning pipeline that includes i) denoising hard negatives, which corrects mislabeling, and ii) large batch training. While this is very effective, the entire pipeline is very heavy in computation and not feasible for people who do not have tremendous hardware resources, especially those in academia. In this paper, we ask, instead of directly using the pipeline, can we take the insights of RocketQA to perform language model pre-training such that the pre-trained model can be easily fine-tuned on any target query set. 
 
 Concretely, we ask what the optimized training in RocketQA solves. We hypothesize that typical LMs are sensitive to mislabeling, which can cause detrimental updates to the model weights. Denoising can effectively remove the bad samples and their updates. On the other hand, for most LMs, the CLS vectors are either trained with a simple task~\cite{devlin-etal-2019-bert} or not explicitly trained at all~\cite{Liu2019RoBERTaAR}. These vectors are far from being able to form an embedding space of passages~\cite{lee-etal-2019-latent}. The large training batches in RocketQA help the LM to stably learn to form the full embedding space. To this end, we want to pre-train an LM such that it is locally noise-resistant and has a well-structured global embedding space. For noise resistance, we borrow the Condenser pre-training architecture~\cite{condenser}, which performs language model pre-training actively conditioned on the CLS vector. It produces an information-rich CLS representation that can robustly condense an input sequence. We then introduce a simple corpus level contrastive learning objective: given a target corpus of documents to retrieve from, at each training step sample text span pairs from a batch of documents and train the model such that the CLS embeddings of two spans from the same document are close and spans from different documents are far apart. Combining the two, we propose coCondenser pre-training, which unsupervisedly learns a corpus-aware pre-trained model for dense retrieval.
 
 In this paper, we test coCondenser pre-training on two popular corpora, Wikipedia and MS-MARCO. Both have served as information sources for a wide range of tasks. This popularity justifies pre-training models specifically for each of them. We directly fine-tune the pre-trained coCondenser using small training batches without data engineering. On Natural Question, TriviaQA, and MS-MARCO passage ranking tasks, we found that the resulting models perform on-par or better than RocketQA and other contemporary methods.
 
\section{Related Work}
\paragraph{Dense Retrieval} Transformer LM has advanced the state-of-the-art of many NLP tasks~\cite{devlin-etal-2019-bert,Liu2019RoBERTaAR,Yang2019XLNetGA,Lan2020ALBERTAL} including dense retrieval. \citet{lee-etal-2019-latent} are among the first to demonstrate the effectiveness of Transformer dense retrievers. They proposed a simple Inverse Cloze Task (ICT) method to further pre-train BERT~\cite{devlin-etal-2019-bert}. Follow-up works explored other pre-training tasks~\cite{chang2020pretraining} as well end-to-end co-training of reader and retriever~\cite{Guu2020REALMRL}. \citet{karpukhin-etal-2020-dense} is the first to discover that careful fine-tuning can learn effective dense retriever directly from BERT. Later works then started to investigate ways to further improve fine-tuning~\cite{xiong2021approximate,Qu2020RocketQAAO}. Among them, \citet{Qu2020RocketQAAO} proposed the RocketQA fine-tuning pipeline which hugely advanced the performance of dense retrievers.  

Until the very recent, pre-training for dense retrieval has been left unexplored. A concurrent work DPR-PAQ~\cite{oguz2021domainmatched} revisits pre-training and prosposes domain matched pre-training, where they use a 65-million-size synthetic QA pair dataset generated with pre-trained Natural Question and Trivia QA pipelines to pre-train dense retrievers. 

In this paper, we will be using a recently proposed dense retrieval pre-training architecture, Condenser~\cite{condenser}. Unlike previous works that design pre-training tasks, Condenser explored the idea of designing special pre-training architecture. 

One reason why dense retrieval is of immediate great value is that there exists rich literature that studies efficient dense retrieval for first stage retrieval~\cite{JDH17,avq_2020}. There exists matured dense retrieval libraries like FAISS~\cite{JDH17}. By pre-encoding the corpus into MIPS index, retrieval can run online with millisecond-level latency~\cite{JDH17,avq_2020}.

\paragraph{Contrastive Learning}
Contrastive learning have become a very popular topic in computer vision~\cite{Chen2020ASF, He2020MomentumCF}. Recent works have brought the idea to natural language processing to learn high-quality sentence representation~\cite{Giorgi2020DeCLUTRDC, Wu2020CLEARCL}. In this work, we use contrastive learning to do pre-training for dense retrieval. Different from earlier work, instead of single representations, we are interested in the full learned embedding space, which we will use to warm start the retriever's embedding space.

The large batch requirement had been a limiting factor in contrastive learning~\cite{Chen2020ASF} and in general any training procedure that uses contrastive loss including dense retrieval pre-training~\cite{Guu2020REALMRL,chang2020pretraining}, for resource-limited setups where GPU memory is not sufficiently large. \citet{gao-etal-2021-scaling} recently devised a gradient cache technique that thresholds peak memory usage of contrastive learning to almost constant. In \autoref{sec:method-cache}, we show how to adapt it for coCondenser pre-training. 
\section{Method}
\label{sec:method}
In this section, we first give a brief overview of Condenser. Then we discuss how to extend it to coCondenser and how to perform memory-efficient coCondenser pre-training.

\begin{figure}[h]
  \centering
  \includegraphics[width=0.47\textwidth]{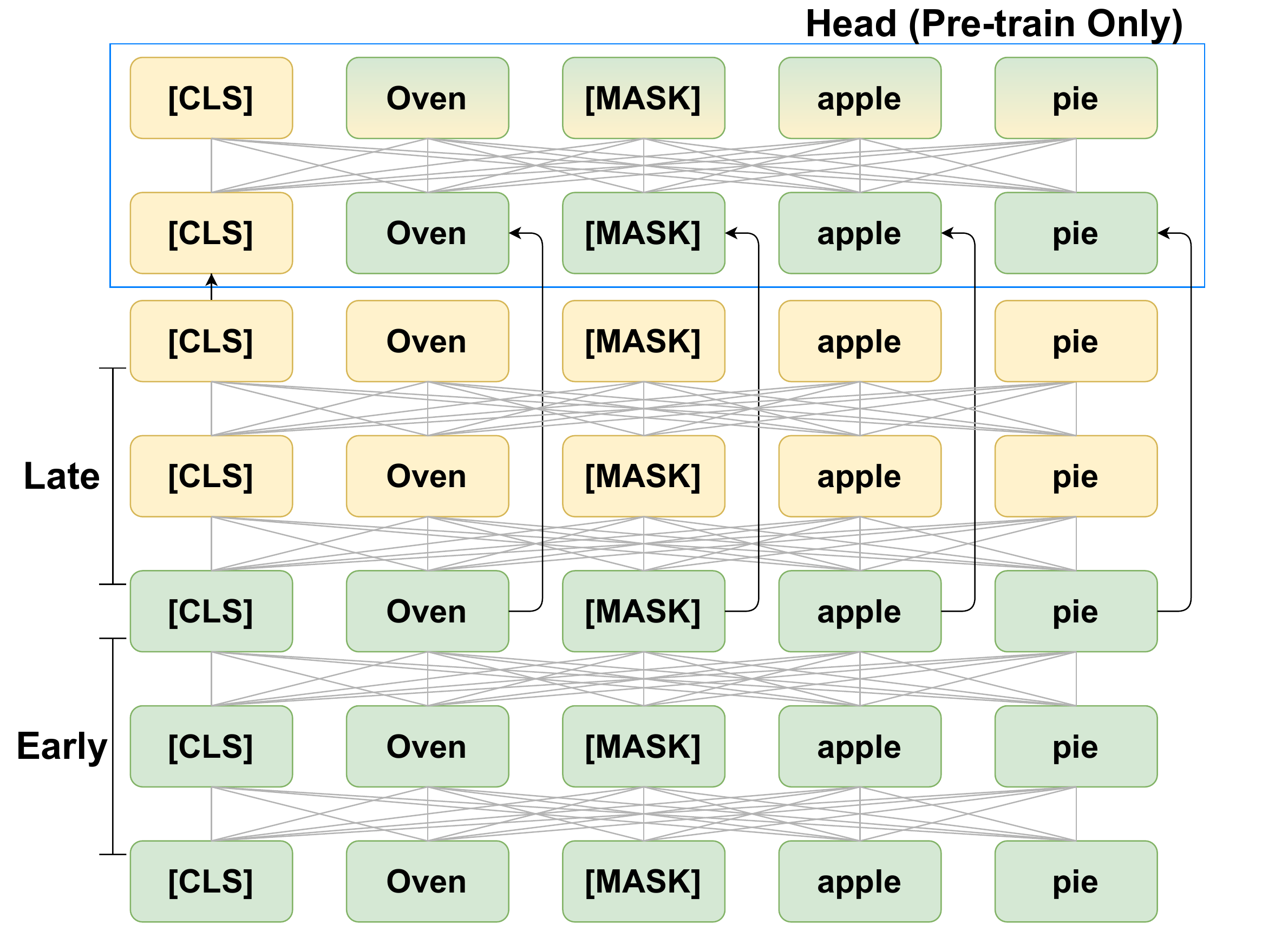}
  \vspace{-2mm}
  \caption{Condenser: Shown are 2 early and 2 late backbone layers.  Our experiments each have 6 layers. Condenser Head is dropped during fine-tuning.}
  \label{fig: model}
\end{figure}

\subsection{Condenser}
\label{subsec:condenser}
In this paper, we adopt a special pre-training architecture Condenser~\cite{condenser}. Condenser is a stack of Transformer blocks. As shown in \autoref{fig: model}, these Transformer blocks are divided into three groups, early backbone encoder layers, late backbone encoder layers, and head layers. An input $x=[x_1, x_2, ..]$ is first prepended a CLS, embedded, and run through the backbone layers.
\begin{align}
    &[h_{cls}^0;h^0] = \text{Embed}([\text{CLS}; x])\\
    &[h_{cls}^{early};h^{early}] = \text{Encoder}_\text{early}([h_{cls}^0;h^0]) \label{eq:cd-input} \\ 
    &[h_{cls}^{late};h^{late}] = \text{Encoder}_\text{late}([h_{cls}^{early};h^{early}]) 
\end{align}
The head takes the CLS representation from the late layers but using a short circuit, the token representations from the early layers. This late-early pair then runs through the head's Transformer blocks.
\begin{equation}
    [h_{cls}^{cd};h^{cd}] = \text{Head}([h_{cls}^{late};h^{early}])
\end{equation}
The head's outputs are then used to perform masked language model~(MLM; \citet{devlin-etal-2019-bert}) training.
\begin{equation}
    \mathcal{L}^\text{mlm} = \sum_{i \in \text{masked}} \text{CrossEntropy}(W h^{cd}_i, x_i)
\end{equation}
To utilize the capacity of the late layers, Condenser is forced to learn to aggregate information into the CLS, which will then participate in the LM prediction. Leveraging the rich and effective training signal produced by MLM, Condenser learn to utilize the powerful Transformer architecture to generate dense CLS representation. We hypothesize that with this LM objective typically used to train token representation now put on the dense CLS representation, the learned LM gains improved robustness against noise.

\subsection{coCondenser}
\label{sec:method-co-condenser}
While Condenser can be trained on a diverse collection of corpra to produce a universal model, it is not able to solve the embedding space issue: while information embedded in the CLS can be non-linearly interpreted by the head, inner products between these vectors still lack semantics. Consequently, they do not form an effective embedding space. To this end, we augment the Condenser MLM loss with a contrastive loss. Unlike previous work that pre-trains on artificial query passage pairs, in this paper, we propose to simply pre-train the passage embedding space in a query-agnostic fashion, using a contrastive loss defined over the target search corpus. Concretely, given a random list of $n$ documents $[d_1, d_2, ..., d_n]$, we extract randomly from each a pair of spans, $[s_{11}, s_{12}, ..., s_{n1}, s_{n2}]$. These spans then form a training batch of coCondenser. Write a span $s_{ij}$'s corresponding \emph{late} CLS representation $h_{ij}$, its corpus-aware contrastive loss is defined over the batch,
\begin{equation}
\label{eq:co-loss}
    \mathcal{L}^{co}_{ij}= - \log \frac{\exp(\langle h_{i1}, h_{i2} \rangle)}{\sum_{k=1}^n \sum_{l=1}^2 \mathbb{I}_{ij \neq kl} \exp(\langle h_{ij}, h_{kl} \rangle) }
\end{equation}
Familiar readers may recognize this as the contrastive loss from SimCLR~\cite{Chen2020ASF}, for which we use random span sampling as augmentation. 
Others may see a connection to noise contrastive estimation~(NCE). Here we provide an NCE narrative. Following the spirit of the distributional hypothesis, passages close together should have similar representations while those in different documents should have different representations. Here we use random spans as surrogates of passages and enforce the distributional hypothesis through NCE, as word embedding learning in Word2Vec~\cite{Mikolov2013EfficientEO}. We can also recognize this as a span-level language model objective, or ``skip-span''. Denote span $s_{ij}$'s Condenser MLM loss $\mathcal{L}^\text{mlm}_{ij}$, the batch's loss is defined as an average sum of MLM and contrastive loss, or from an alternative perspective, word and span LM loss,

\begin{equation}
\label{eq:full-loss}
    \mathcal{L} = \frac{1}{2n} \sum_{i=1}^{n} \sum_{j=1}^2 [\mathcal{L}^\text{mlm}_{ij} + \mathcal{L}_{ij}^{co}]
\end{equation}

\subsection{Memory Efficient Pre-training}
\label{sec:method-cache}
The RocketQA pipeline uses supervision and large-batch training to learn the embedding space. We would also like to run large-batch unsupervised pre-training to construct effective stochastic gradient estimators for the contrastive loss in \autoref{eq:co-loss}. To remind our readers, this large-batch pre-training happens only once for the target search corpus. We will show that this allows effective small batch fine-tuning on task query sets.

However, due to the batch-wise dependency of the contrastive loss, it requires fitting the large batch into GPU~(accelerator) memory. While this can done naively with interconnected GPU nodes or TPU pods, which can have thousands of gigabytes of memory, academia and smaller organizations are often restricted to machines with four commercial GPUs. To break the memory constraint and perform effective contrastive learning, we incorporate the gradient caching technique~\cite{gao-etal-2021-scaling}. We describe the procedure here for people that want to perform coCondenser pre-training but have limited resources. Denote $\mathcal{L}^{co} = \sum_i \sum_j \mathcal{L}_{ij}^{co}$, we can write \autoref{eq:full-loss} as,
\begin{equation}
\label{eq:full-loss-co}
    \mathcal{L} =\frac{1}{2n} [ \mathcal{L}^{co} +  \sum_i \sum_j \mathcal{L}^\text{mlm}_{ij}]
\end{equation}
The spirit of gradient caching is to decouple representation gradient and encoder gradient computation. Before computing the model weight update, we first run an extra backbone forward for the entire batch, without constructing the computation graph. This provides the numerical values of $[h_{11}, h_{12}, ...., h_{n1}, h_{n2}]$, from which we can compute:
\begin{equation}
    v_{ij} 
    = \frac{\partial }{\partial h_{ij}} \sum_i \sum_j \mathcal{L}_{ij}^{co}
    = \frac{\partial \mathcal{L}^{co}}{\partial h_{ij}}
\end{equation}
i.e. the contrastive loss gradient with respect to the CLS vector. We store all these vectors in a gradient cache, $C = [v_{11}, v_{12}, .., v_{n1}, v_{n2}]$. Using $v_{ij}$, denote the model parameter $\Theta$, we can write the derivative of the contrastive loss as shown below.
\begin{align}
     \frac{\partial \mathcal{L}^{co}}{\partial \Theta}
     &= \sum_i \sum_j \frac{\partial \mathcal{L}^{co} }{\partial h_{ij}} \frac{\partial  h_{ij}}{\partial \Theta}\\
     &= \sum_i \sum_j v_{ij}^{\top} \frac{\partial  h_{ij}}{\partial \Theta} 
\end{align}
We can then write the gradient of \autoref{eq:full-loss-co}.
\begin{align}
    \frac{\partial \mathcal{L}}{\partial \Theta} = \frac{1}{2n} \sum_i \sum_j [v_{ij}^{\top} \frac{\partial  h_{ij}}{\partial \Theta} + \frac{\partial\mathcal{L}^\text{mlm}_{ij}}{\partial \Theta} ]
\end{align}
Since $v_{ij}$ is already in the cache $C$, each summation term now only concerns span $s_{ij}$ and its activation, meaning that we can compute the full batch's gradient in an accumulation fashion over small sub-batches. In other words, the full batch no longer needs to concurrently reside on the GPUs.

\begin{figure*}[h]
  \centering
  \begin{subfigure}[b]{0.97\textwidth}
  \includegraphics[width=0.99\textwidth]{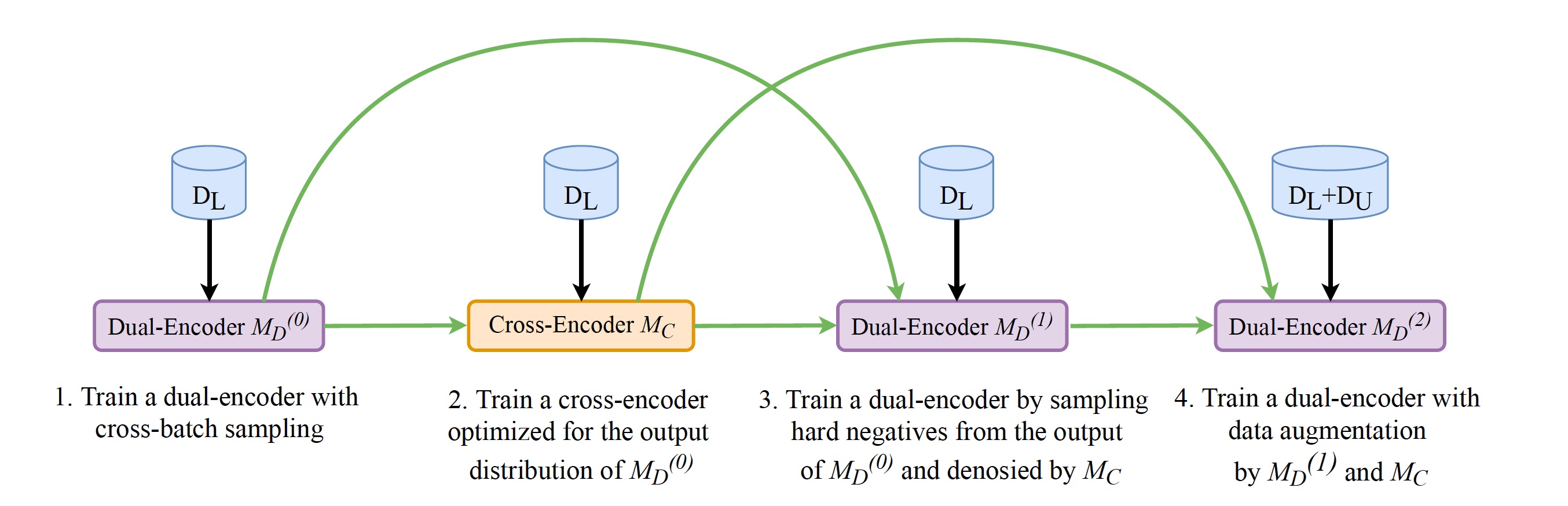}
  \caption{RocketQA retriever training pipeline (taken from \citet{Qu2020RocketQAAO}).}
  \label{fig:rocket-pipeline}
  \end{subfigure}
  \begin{subfigure}[b]{0.55\textwidth}
  \includegraphics[width=0.99\textwidth]{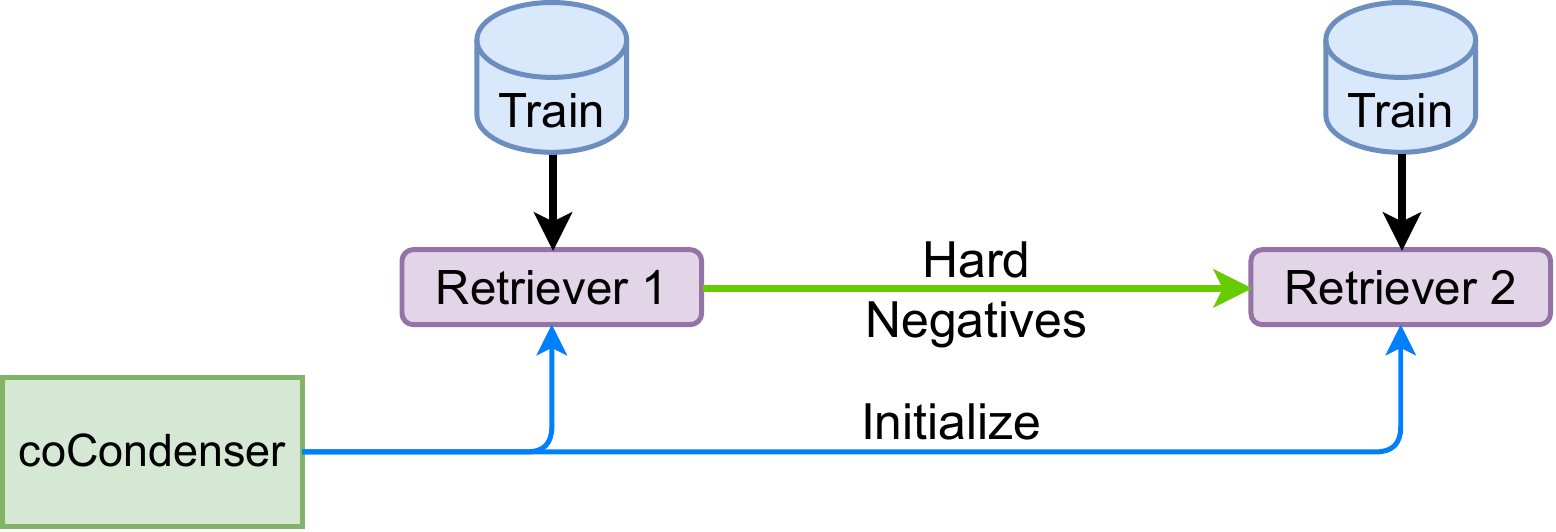}
  \caption{coCondenser retriever training pipeline.}
  \label{fig:co-pipeline}
  \end{subfigure}
  \caption{RocketQA training pipelines and two-round retriever training pipeline in coCondenser.}
  \label{fig:pipelines}
\end{figure*}

\subsection{Fine-tuning}
\label{sec:method-fine-tune}
At the end of pre-training, we discard the Condenser head, keeping only the backbone layers. Consequently, the model reduces to its backbone, or effectively a Transformer Encoder. We use the backbone weights to initialize query encoder $f_q$ and passage encoder $f_p$, each outputs the last layer CLS. Recall that they have already been warmed up in pre-training. A (query $q$, passage $p$) pair similarity is defined as an inner product,
\begin{equation}
    s(q, p) = \langle f_q(q), f_p(p) \rangle
\end{equation}
Query and passage encoders are supervisedly fine-tuned on the target task's training set. We train with a supervised contrastive loss and compute for query $q$, negative log likelihood of a positive document $d^+$ against a set of negatives $\{d^-_1, d^-_2, .. d^-_l ..\}$. 
\begin{equation}
\mathcal{L} = -\log \frac{\exp(s(q, d^+))}{ \exp(s(q, d^+)) + \underset{l}{\sum} \exp(s(q, d^-_l)) }
\end{equation}
We run a two-round training as described in the DPR~\cite{karpukhin-etal-2020-dense} toolkit. As shown in \autoref{fig:co-pipeline}, in the first round, the retrievers are trained with BM25 negatives. The first-round retriever is then used to mine hard negatives to complement the negative pool. The second round retriever trains with the negative pool generated in the first round. This is in contrast to the multi-stage pipeline of RocketQA shown in \autoref{fig:rocket-pipeline}. 
\section{Experiments}
\label{sec:experiment}
\begin{table*}[t]
\centering
\scalebox{0.95}{
\begin{tabular}{ l c c  c c c c c c}
\hline 
\multirow{ 2}{*}{Method}& \multicolumn{2}{c}{\textbf{MS-MARCO Dev}} & \multicolumn{3}{c}{\textbf{Natural Question Test}} & \multicolumn{3}{c}{\textbf{Trivia QA Test}}  \\
& MRR@10 & R@1000 & R@5 & R@20 & R@100 & R@5 & R@20 & R@100 \\
\hline
BM25 & 18.7 & 85.7 & - & 59.1 & 73.7 & - & 66.9 & 76.7 \\
DeepCT & 24.3 & 90.9 & - & - & - & - & - & -\\
docT5query & 27.7 & 94.7 & - & - & - & - & - & -  \\
GAR & - & - & 60.9 & 74.4 & 85.3 & 73.1 & 80.4 & 85.7 \\
\hline
DPR & - & - & - & 74.4 & 85.3 & - & 79.3 & 84.9 \\
ANCE & 33.0 & 95.9 & - & 81.9 & 87.5 & - & 80.3 & 85.3\\
ME-BERT & 33.8 & -  & - & - & - & - & - & - \\
\hline
RocketQA & 37.0 & 97.9 & 74.0 & 82.7 & 88.5 & - & - & - \\
\hline
Condenser & 36.6 & 97.4 & - & 83.2 & 88.4 & - & 81.9 & 86.2\\

\hline
DPR-PAQ \\
- BERT$_\text{base}$ & 31.4 & - & 74.5 & 83.7 & 88.6 & - & - & - \\
- BERT$_\text{large}$ & 31.1 & - & 75.3 & 84.4 & 88.9 & - & - & -\\
- RoBERTa$_\text{base}$ & 32.3 & - & 74.2 & 84.0 & \textbf{89.2} & - & - & -\\
- RoBERTa$_\text{large}$ & 34.0 & - & \textbf{76.9} & \textbf{84.7} & \textbf{89.2} & - & - & -\\
\hline
coCondenser & \textbf{38.2} & \textbf{98.4} & \textbf{75.8} & \textbf{84.3} & 89.0 & \textbf{76.8} & \textbf{83.2} & \textbf{87.3}\\
\hline 
\end{tabular}
}
\caption{Retrieval performance on MSMARCO dev, Natural Question test and Trivia QA test. We mark bold the best performing models as well as the best performing 12-layer base models. Results unavailable are left blank.}
\label{tab:results}
\end{table*}
In this section, we first describe the implementation details of coCondenser pre-training. We then conduct dense retrieval experiments to test the effectiveness of fine-tuned coCondenser retrievers.

\subsection{Pre-training}
The coCondenser pre-training starts with vanilla BERT and goes in two stages, universal Condenser pre-training and corpus aware coCondenser pre-training. In the first stage, we pre-train a Condenser and warm start the backbone layers with pre-trained 12-layer BERT$_\text{base}$ weights~\cite{devlin-etal-2019-bert}. The backbone uses an equal split, 6 early layers, and 6 late layers. The Condenser pre-training uses the same data as BERT: English Wikipedia and the BookCorpus. The Condenser model from stage one, including both backbone and head, is taken to warm start stage two coCondenser pre-training on the target corpus~(Wikipedia or MS-MARCO web collection). We keep the Condenser architecture unchanged in the second step. We use AdamW optimizer with a learning rate 1e-4, weight decay of 0.01, and linear learning rate decay. Each model weight update uses 2K documents. We train on 4 RTX 2080 Ti GPUs using gradient cache update, as described in \autoref{sec:method-cache}.

After the second step finishes, we discard the Condenser head, resulting in a model of the \emph{exact} same architecture as BERT$_\text{base}$.

\subsection{Dense Passage Retrieval}
Next, we fine-tune the learned coCondenser to test retrieval performance. Following RocketQA, we test on Natural Question and MS-MARCO passage ranking. We also report performance on Trivia QA, whose pre-processed version is released with the DPR toolkit.

\subsubsection{Setup}
\paragraph{Dataset} We use MS-MARCO passage ranking~\cite{bajaj2018ms},  Natural Question(NQ; \citet{kwiatkowski-etal-2019-natural}) and Trivia QA(TQA; \citet{joshi-etal-2017-triviaqa}). MS-MARCO is constructed from Bing's search query logs and web documents retrieved by Bing. Natural Question contains questions from Google search. Trivia QA contains a set of trivia questions. We report official metrics MRR@10, Recall@1000 for MS-MARCO, and Recall at 5, 20, and 100 for NQ and TQA.

\paragraph{Data Preparation} We use Natural Question, Trivia QA, and Wikipedia as cleaned and released with DPR toolkit. NQ and TQA have about 60K training data post-processing. Similarly, we use the MS-MARCO corpus released with RocketQA open-source code. For reproducibility, we use the official relevance file instead of RocketQA's extended one, which has about 0.5M training queries. The BM25 negatives for MS-MARCO are taken from the official training triples. 

\paragraph{Training} MS-MARCO models are trained using AdamW with a 5e-6 learning rate, linear learning rate schedule, and batch size 64 for $3$ epochs. Models are trained only on each task's corresponding training. We note that RocketQA is trained on a concatenation of several datasets~\cite{Qu2020RocketQAAO}. NQ and TQA models are trained with the DPR toolkit following published hyperparameters by \citet{karpukhin-etal-2020-dense}. All models are trained on one RTX 2080 Ti. We added gradient caching to DPR to deal with memory constraints\footnote{\url{https://github.com/luyug/GC-DPR}}.

\paragraph{Model Validation} Since for dense retrieval, validating a checkpoint requires encoding the full corpus, evaluating a checkpoint becomes very costly. Due to our computation resource limitation, we follow the suggestion in the DPR toolkit and take the last model training checkpoint. We do the same for MS-MARCO.

\paragraph{Comparison Systems} We used RocketQA~\cite{Qu2020RocketQAAO}, the state-of-the-art fine-tuning technique, as our main baseline. 

We borrowed several other baselines from the RocketQA paper, including lexical systems BM25, DeepCT~\cite{DeepCT}, DocT5Query~\cite{docTTTTTquery} and GAR~\cite{mao2020generationaugmented}; and dense systems DPR~\cite{karpukhin-etal-2020-dense}, ANCE~\cite{xiong2021approximate}, and ME-BERT~\cite{Luan2020SparseDA}. 

We also included the concurrent work DPR-PAQ~\cite{oguz2021domainmatched}, which pre-trains using a 65-million-size synthetic QA pair dataset. The pre-training data is created by using retriever-reader pairs trained on Natural Question and Trivia QA. Designing the synthesis procedure also requires domain knowledge, thus we refer to this as a \emph{semi-supervised} pre-training method. We include 4 DPR-PAQ variants based on base/large architectures of BERT/RoBERTa models.

Finally, we fine-tune a Condenser model which is produced in the first stage of pre-training.

\begin{table*}[t]
\centering
\scalebox{0.95}{
\begin{tabular}{ l  c c c}
\hline 
\multirow{ 2}{*}{\textbf{Method}} & \multirow{ 2}{*}{\textbf{Batch Size}} & \multicolumn{2}{c}{\textbf{MS-MARCO Dev}}   \\
&   & MRR@10 & R@1000  \\
\hline
\textbf{RocketQA}\\
Cross-batch negatives  & 8192 & 33.3 & -  \\
+ Hard negatives  & 4096 & 26.0 & -  \\
+ Denoising  & 4096 & 36.4 & -  \\
+ Data augmentation  & 4096 & 37.0 & 97.9  \\
\hline
\textbf{coCondenser}\\
Condenser w/o Hard negatives  & 64 & 33.8 & 96.1 \\
+ Hard negatives   & 64 & 36.6 & 97.4  \\
coCondenser  w/o Hard negatives  & 64 & 35.7 & 97.8 \\
+ Hard negatives  & 64 & \textbf{38.2} & \textbf{98.4}  \\
\hline 
\end{tabular}
}
\caption{Retrieval performance on the MS-MARCO development (dev) set for various fine-tuning stages of RocketQA and various pre-training and fine-tuning stages of coCondenser.}
\label{tab:stages}
\end{table*}

\subsubsection{Results} \autoref{tab:results} shows development (dev) set performance for MS-MARCO passage ranking and test set performance for Natural Question and Trivia QA. Across three query sets, dense systems show superior performance compared to sparse systems. We also see a big performance margin between systems involving either careful fine-tuning or pre-training (RocketQA, DPR-PAQ, Condenser, coCondenser) over earlier dense systems. This result confirms recent findings that low dimension embeddings possess a strong capacity for dense retrieval, a capacity however hard to exploit naively. 

coCondenser shows small improvements over RocketQA. Importantly, this is achieved with \emph{greatly} reduced computation and data engineering effort in fine-tuning. Notably on MS-MARCO, coCondenser reduced the RocketQA's 4096 batch size to 64~(\autoref{tab:stages}). A comparison of the two training pipelines of RocketQA and coCondenser can be found in \autoref{fig:pipelines}.

Comparison with DPR-PAQ shows several interesting findings. Combining large semi-supervised pre-training with the better and larger LM RoBERTa$_\text{large}$, DPR-PAQ achieves the best results on Natural Question. On the other hand, when starting from BERT~(base/large), DPR-PAQ show similar performance to coCondenser, which is based on BERT$_\text{base}$. This suggests that large-scale semi-supervised pre-training is still the way to go to get the very best performance. However, when computational resources are limited and a large pre-training set is missing, the unsupervised coCondenser is a strong alternative. On the other hand, as it moves to MS-MARCO where DPR-PAQ's pre-training supervision becomes distant, we observe that DPR-PAQ becomes less effective than RocketQA and coCondenser. 

The comparison between Condenser and coCondenser demonstrates the importance of the contrastive loss in coCondener: coCondenser can be robustly fine-tuned thanks to its pre-structured embedding space, allowing it to have better Recall (fewer false negatives) across all datasets. 

\subsection{Passage Reranking on MS-MARCO}
We also tested reranking coCondenser results with a deep LM reranker. Similar to \citet{Qu2020RocketQAAO}, we train an ensemble of ERNIE and RoBERTa to rerank the top 1000 retrieved passages on the MS-MARCO evaluation set and test them on Microsoft's
hidden test set. \autoref{tab:lb} shows the top three systems on August 11, 2021. 

coCondenser is best by a small, perhaps insignificant margin. Essentially, the three systems represent three distinct and equally good approaches for effective web passage retrieval: optimized dense retrieval fine-tuning in RocketQA~\cite{Qu2020RocketQAAO}, contextualized sparse retrieval in COIL~\cite{gao-etal-2021-coil}, and corpus-aware unsupervised pre-training in coCondenser.

\begin{table}
\centering
\scalebox{0.9}{
\begin{tabular}{c l c}
\hline
\multirow{ 2}{*}{\textbf{Rank}} & \multirow{ 2}{*}{\textbf{Method}}  & \textbf{EVAL}   \\
& & \textbf{MRR@10}\\
\hline
1 & coCondenser & 42.8 \\
2 & C-COIL~\cite{gao-etal-2021-coil} & 42.7 \\
3 & RocketQA & 42.6 \\
\hline
\end{tabular}
}
\vspace{-2mm}
\caption{Reranking performance on the MS-MARCO passage ranking leaderboard.}
\label{tab:lb}
\vspace{-3mm}
\end{table}

\section{Analysis of Training Stages}
\label{sec:analysis}
Next, we seek to understand the contribution of each pre-training and fine-tuning stage of coCondenser retriever. We consider pre-trained Condenser from the first stage and coCondenser from the second stage. For each, we consider retrievers trained with and without hard negatives.  For reference, we compare with various RocketQA training stages. Results are shown in \autoref{tab:stages}. 

We see that each stage of RocketQA is critical.  As each is added, performance improves steadily. On the other hand, this also suggests the full pipeline has to be executed to get the best performance. 

In comparison, we see Condenser with hard negatives has performance very close to the full RocketQA system. Condenser with hard negatives also has better MRR than coCondenser without hard negatives, meaning that Condenser from the first pre-training stage is already very strong locally but the embedding space trained from a relatively cold start is still not optimal, causing global misses. 

Adding the corpus aware loss, coCondenser without hard negatives has Recall very close to the full RocketQA system, using only a size 64 batch. This confirms our hypothesis that fine-tuning can benefit from a pre-trained passage embedding space. Further adding hard negatives, we get the strongest coCondenser system that is both locally and globally effective.  Note that all Condenser systems achieve their performance \emph{without} denoising, showing the superior noise resistance capability learned using the Condenser architecture. Practically, our systems also do not require data augmentation,  which helps reduce engineering effort in designing augmentation techniques and defining augmentation data.

To summarize, the coCondenser pre-training has achieved the goals we set for it. It can be effectively fine-tuned without relying on the RocketQA techniques: denoise hard negatives, large training batch, or data augmentation, simply using mined hard negatives with small training batches. 

\section{Conclusion}
This paper introduces \emph{coCondenser}, an unsupervised corpus-aware language model pre-training method. 
Leveraging the Condenser architecture and a corpus aware contrastive loss, coCondenser acquires two important properties for dense retrieval, noise resistance and structured embedding space. This corpus-aware pre-training needs to be done once for a search corpus and is query agnostic. The learned model can be shared among various types of end task queries.  

Experiments show that coCondenser can drastically reduce the costs of fine-tuning a dense retriever while also improving retrieval performance. They also show that coCondenser yields performance close or similar to models that are several times larger and require semi-supervised pre-training. 

Importantly, coCondenser provides a completely \emph{hands-off} way to pre-train a very effective LM for dense retrieval. This effectively removes the effort for designing and testing pre-training as well as fine-tuning techniques. For practitioners, by adopting our pre-trained weight, they can use limited resource to train dense retrieval systems with state-of-the-art performance. 

On the other hand, future works may also investigate integrating extra well-tested pre-training/fine-tuning methods to further improve performance. 
\newpage

\bibliography{anthology,acl2021}

\begin{thebibliography}{28}
\expandafter\ifx\csname natexlab\endcsname\relax\def\natexlab#1{#1}\fi

\bibitem[{Bajaj et~al.(2018)Bajaj, Campos, Craswell, Deng, Gao, Liu, Majumder,
  McNamara, Mitra, Nguyen, Rosenberg, Song, Stoica, Tiwary, and
  Wang}]{bajaj2018ms}
Payal Bajaj, Daniel Campos, Nick Craswell, Li~Deng, Jianfeng Gao, Xiaodong Liu,
  Rangan Majumder, Andrew McNamara, Bhaskar Mitra, Tri Nguyen, Mir Rosenberg,
  Xia Song, Alina Stoica, Saurabh Tiwary, and Tong Wang. 2018.
\newblock \href {http://arxiv.org/abs/1611.09268} {Ms marco: A human generated
  machine reading comprehension dataset}.

\bibitem[{Chang et~al.(2020)Chang, Yu, Chang, Yang, and
  Kumar}]{chang2020pretraining}
Wei-Cheng Chang, Felix~X. Yu, Yin-Wen Chang, Yiming Yang, and Sanjiv Kumar.
  2020.
\newblock \href {https://openreview.net/forum?id=rkg-mA4FDr} {Pre-training
  tasks for embedding-based large-scale retrieval}.
\newblock In \emph{International Conference on Learning Representations}.

\bibitem[{Chen et~al.(2020)Chen, Kornblith, Norouzi, and Hinton}]{Chen2020ASF}
Ting Chen, Simon Kornblith, Mohammad Norouzi, and Geoffrey~E. Hinton. 2020.
\newblock A simple framework for contrastive learning of visual
  representations.
\newblock \emph{ArXiv}, abs/2002.05709.

\bibitem[{Dai and Callan(2019)}]{DeepCT}
Zhuyun Dai and J.~Callan. 2019.
\newblock Context-aware sentence/passage term importance estimation for first
  stage retrieval.
\newblock \emph{ArXiv}, abs/1910.10687.

\bibitem[{Devlin et~al.(2019)Devlin, Chang, Lee, and
  Toutanova}]{devlin-etal-2019-bert}
Jacob Devlin, Ming-Wei Chang, Kenton Lee, and Kristina Toutanova. 2019.
\newblock \href {https://doi.org/10.18653/v1/N19-1423} {{BERT}: Pre-training of
  deep bidirectional transformers for language understanding}.
\newblock In \emph{Proceedings of the 2019 Conference of the North {A}merican
  Chapter of the Association for Computational Linguistics: Human Language
  Technologies, Volume 1 (Long and Short Papers)}, pages 4171--4186,
  Minneapolis, Minnesota. Association for Computational Linguistics.

\bibitem[{Gao and Callan(2021)}]{condenser}
Luyu Gao and Jamie Callan. 2021.
\newblock \href {http://arxiv.org/abs/2104.08253} {Is your language model ready
  for dense representation fine-tuning?}

\bibitem[{Gao et~al.(2021{\natexlab{a}})Gao, Dai, and
  Callan}]{gao-etal-2021-coil}
Luyu Gao, Zhuyun Dai, and Jamie Callan. 2021{\natexlab{a}}.
\newblock \href {https://doi.org/10.18653/v1/2021.naacl-main.241} {{COIL}:
  Revisit exact lexical match in information retrieval with contextualized
  inverted list}.
\newblock In \emph{Proceedings of the 2021 Conference of the North American
  Chapter of the Association for Computational Linguistics: Human Language
  Technologies}, pages 3030--3042, Online. Association for Computational
  Linguistics.

\bibitem[{Gao et~al.(2021{\natexlab{b}})Gao, Zhang, Han, and
  Callan}]{gao-etal-2021-scaling}
Luyu Gao, Yunyi Zhang, Jiawei Han, and Jamie Callan. 2021{\natexlab{b}}.
\newblock \href {https://doi.org/10.18653/v1/2021.repl4nlp-1.31} {Scaling deep
  contrastive learning batch size under memory limited setup}.
\newblock In \emph{Proceedings of the 6th Workshop on Representation Learning
  for NLP (RepL4NLP-2021)}, pages 316--321, Online. Association for
  Computational Linguistics.

\bibitem[{Giorgi et~al.(2020)Giorgi, Nitski, Bader, and
  Wang}]{Giorgi2020DeCLUTRDC}
John~Michael Giorgi, Osvald Nitski, Gary~D Bader, and Bo~Wang. 2020.
\newblock Declutr: Deep contrastive learning for unsupervised textual
  representations.
\newblock \emph{ArXiv}, abs/2006.03659.

\bibitem[{Guo et~al.(2020)Guo, Sun, Lindgren, Geng, Simcha, Chern, and
  Kumar}]{avq_2020}
Ruiqi Guo, Philip Sun, Erik Lindgren, Quan Geng, David Simcha, Felix Chern, and
  Sanjiv Kumar. 2020.
\newblock \href {https://arxiv.org/abs/1908.10396} {Accelerating large-scale
  inference with anisotropic vector quantization}.
\newblock In \emph{International Conference on Machine Learning}.

\bibitem[{Guu et~al.(2020)Guu, Lee, Tung, Pasupat, and Chang}]{Guu2020REALMRL}
Kelvin Guu, Kenton Lee, Z.~Tung, Panupong Pasupat, and Ming-Wei Chang. 2020.
\newblock Realm: Retrieval-augmented language model pre-training.
\newblock \emph{ArXiv}, abs/2002.08909.

\bibitem[{He et~al.(2020)He, Fan, Wu, Xie, and Girshick}]{He2020MomentumCF}
Kaiming He, Haoqi Fan, Yuxin Wu, Saining Xie, and Ross~B. Girshick. 2020.
\newblock Momentum contrast for unsupervised visual representation learning.
\newblock \emph{2020 IEEE/CVF Conference on Computer Vision and Pattern
  Recognition (CVPR)}, pages 9726--9735.

\bibitem[{Johnson et~al.(2017)Johnson, Douze, and J{\'e}gou}]{JDH17}
Jeff Johnson, Matthijs Douze, and Herv{\'e} J{\'e}gou. 2017.
\newblock Billion-scale similarity search with gpus.
\newblock \emph{arXiv preprint arXiv:1702.08734}.

\bibitem[{Joshi et~al.(2017)Joshi, Choi, Weld, and
  Zettlemoyer}]{joshi-etal-2017-triviaqa}
Mandar Joshi, Eunsol Choi, Daniel Weld, and Luke Zettlemoyer. 2017.
\newblock \href {https://doi.org/10.18653/v1/P17-1147} {{T}rivia{QA}: A large
  scale distantly supervised challenge dataset for reading comprehension}.
\newblock In \emph{Proceedings of the 55th Annual Meeting of the Association
  for Computational Linguistics (Volume 1: Long Papers)}, pages 1601--1611,
  Vancouver, Canada. Association for Computational Linguistics.

\bibitem[{Karpukhin et~al.(2020)Karpukhin, Oguz, Min, Lewis, Wu, Edunov, Chen,
  and Yih}]{karpukhin-etal-2020-dense}
Vladimir Karpukhin, Barlas Oguz, Sewon Min, Patrick Lewis, Ledell Wu, Sergey
  Edunov, Danqi Chen, and Wen-tau Yih. 2020.
\newblock \href {https://doi.org/10.18653/v1/2020.emnlp-main.550} {Dense
  passage retrieval for open-domain question answering}.
\newblock In \emph{Proceedings of the 2020 Conference on Empirical Methods in
  Natural Language Processing (EMNLP)}, pages 6769--6781, Online. Association
  for Computational Linguistics.

\bibitem[{Kwiatkowski et~al.(2019)Kwiatkowski, Palomaki, Redfield, Collins,
  Parikh, Alberti, Epstein, Polosukhin, Devlin, Lee, Toutanova, Jones, Kelcey,
  Chang, Dai, Uszkoreit, Le, and Petrov}]{kwiatkowski-etal-2019-natural}
Tom Kwiatkowski, Jennimaria Palomaki, Olivia Redfield, Michael Collins, Ankur
  Parikh, Chris Alberti, Danielle Epstein, Illia Polosukhin, Jacob Devlin,
  Kenton Lee, Kristina Toutanova, Llion Jones, Matthew Kelcey, Ming-Wei Chang,
  Andrew~M. Dai, Jakob Uszkoreit, Quoc Le, and Slav Petrov. 2019.
\newblock \href {https://doi.org/10.1162/tacl_a_00276} {Natural questions: A
  benchmark for question answering research}.
\newblock \emph{Transactions of the Association for Computational Linguistics},
  7:452--466.

\bibitem[{Lan et~al.(2020)Lan, Chen, Goodman, Gimpel, Sharma, and
  Soricut}]{Lan2020ALBERTAL}
Zhenzhong Lan, Mingda Chen, Sebastian Goodman, Kevin Gimpel, Piyush Sharma, and
  Radu Soricut. 2020.
\newblock Albert: A lite bert for self-supervised learning of language
  representations.
\newblock \emph{ArXiv}, abs/1909.11942.

\bibitem[{Lee et~al.(2019)Lee, Chang, and Toutanova}]{lee-etal-2019-latent}
Kenton Lee, Ming-Wei Chang, and Kristina Toutanova. 2019.
\newblock \href {https://doi.org/10.18653/v1/P19-1612} {Latent retrieval for
  weakly supervised open domain question answering}.
\newblock In \emph{Proceedings of the 57th Annual Meeting of the Association
  for Computational Linguistics}, pages 6086--6096, Florence, Italy.
  Association for Computational Linguistics.

\bibitem[{Liu et~al.(2019)Liu, Ott, Goyal, Du, Joshi, Chen, Levy, Lewis,
  Zettlemoyer, and Stoyanov}]{Liu2019RoBERTaAR}
Y.~Liu, Myle Ott, Naman Goyal, Jingfei Du, Mandar Joshi, Danqi Chen, Omer Levy,
  M.~Lewis, Luke Zettlemoyer, and Veselin Stoyanov. 2019.
\newblock Roberta: A robustly optimized bert pretraining approach.
\newblock \emph{ArXiv}, abs/1907.11692.

\bibitem[{Luan et~al.(2020)Luan, Eisenstein, Toutanova, and
  Collins}]{Luan2020SparseDA}
Y.~Luan, Jacob Eisenstein, Kristina Toutanova, and Michael Collins. 2020.
\newblock Sparse, dense, and attentional representations for text retrieval.
\newblock \emph{ArXiv}, abs/2005.00181.

\bibitem[{Mao et~al.(2020)Mao, He, Liu, Shen, Gao, Han, and
  Chen}]{mao2020generationaugmented}
Yuning Mao, Pengcheng He, Xiaodong Liu, Yelong Shen, Jianfeng Gao, Jiawei Han,
  and Weizhu Chen. 2020.
\newblock \href {http://arxiv.org/abs/2009.08553} {Generation-augmented
  retrieval for open-domain question answering}.

\bibitem[{Mikolov et~al.(2013)Mikolov, Chen, Corrado, and
  Dean}]{Mikolov2013EfficientEO}
Tomas Mikolov, Kai Chen, G.~S. Corrado, and J.~Dean. 2013.
\newblock Efficient estimation of word representations in vector space.
\newblock In \emph{ICLR}.

\bibitem[{Nogueira and Lin(2019)}]{docTTTTTquery}
Rodrigo Nogueira and Jimmy Lin. 2019.
\newblock From doc2query to doctttttquery.

\bibitem[{Oğuz et~al.(2021)Oğuz, Lakhotia, Gupta, Lewis, Karpukhin, Piktus,
  Chen, Riedel, tau Yih, Gupta, and Mehdad}]{oguz2021domainmatched}
Barlas Oğuz, Kushal Lakhotia, Anchit Gupta, Patrick Lewis, Vladimir Karpukhin,
  Aleksandra Piktus, Xilun Chen, Sebastian Riedel, Wen tau Yih, Sonal Gupta,
  and Yashar Mehdad. 2021.
\newblock \href {http://arxiv.org/abs/2107.13602} {Domain-matched pre-training
  tasks for dense retrieval}.

\bibitem[{Qu et~al.(2021)Qu, Ding, Liu, Liu, Ren, Zhao, Dong, Wu, and
  Wang}]{Qu2020RocketQAAO}
Yingqi Qu, Yuchen Ding, Jing Liu, Kai Liu, Ruiyang Ren, Wayne~Xin Zhao, Daxiang
  Dong, Hua Wu, and Haifeng Wang. 2021.
\newblock \href {https://doi.org/10.18653/v1/2021.naacl-main.466}
  {{R}ocket{QA}: An optimized training approach to dense passage retrieval for
  open-domain question answering}.
\newblock In \emph{Proceedings of the 2021 Conference of the North American
  Chapter of the Association for Computational Linguistics: Human Language
  Technologies}, pages 5835--5847, Online. Association for Computational
  Linguistics.

\bibitem[{Wu et~al.(2020)Wu, Wang, Gu, Khabsa, Sun, and Ma}]{Wu2020CLEARCL}
Z.~Wu, Sinong Wang, Jiatao Gu, Madian Khabsa, Fei Sun, and Hao Ma. 2020.
\newblock Clear: Contrastive learning for sentence representation.
\newblock \emph{ArXiv}, abs/2012.15466.

\bibitem[{Xiong et~al.(2021)Xiong, Xiong, Li, Tang, Liu, Bennett, Ahmed, and
  Overwijk}]{xiong2021approximate}
Lee Xiong, Chenyan Xiong, Ye~Li, Kwok-Fung Tang, Jialin Liu, Paul~N. Bennett,
  Junaid Ahmed, and Arnold Overwijk. 2021.
\newblock \href {https://openreview.net/forum?id=zeFrfgyZln} {Approximate
  nearest neighbor negative contrastive learning for dense text retrieval}.
\newblock In \emph{International Conference on Learning Representations}.

\bibitem[{Yang et~al.(2019)Yang, Dai, Yang, Carbonell, Salakhutdinov, and
  Le}]{Yang2019XLNetGA}
Z.~Yang, Zihang Dai, Yiming Yang, J.~Carbonell, R.~Salakhutdinov, and Quoc~V.
  Le. 2019.
\newblock Xlnet: Generalized autoregressive pretraining for language
  understanding.
\newblock In \emph{NeurIPS}.

\end{thebibliography}
\bibliographystyle{acl_natbib}


\end{document}